\documentclass[secnumarabic,preprint,superscriptaddress,floatfix]{revtex4-2}
\usepackage[english]{babel}
\usepackage{graphicx}
\usepackage{color}
\usepackage{amsmath}
\usepackage{amssymb}
\usepackage{tabularx}
\usepackage[dvipsnames]{xcolor}

\begin{document}
\title{Electronic properties of the mean-field resonating valence bond model of cuprates}
	\author{Sergei Urazhdin}
	\email{sergei.urazhdin@emory.edu}
	\affiliation{Department of Physics, Emory University, Atlanta, USA}
	\author{Sergei Ivanov}
	\affiliation{Department of Physics, Emory University, Atlanta, USA}
	\date{\today}
\begin{abstract} We show that the mean-field resonating valence bond approximation proposed in 1987 by Baskaran, Zou, and Anderson describes gapless charge pair excitations confined to the boundaries of the spinon Brillouin zone. The existence of such pairs accounts for all the essential anomalous electronic properties of cuprates, with superconductivity arising due to the charge drag by the spinon superflow. This mechanism may be relevant to other unconventional superconductors.
\end{abstract}
	
\maketitle

\section{Introduction}\label{sec:intro}

Cuprate high-temperature superconductors (HTSCs) such as Bi$_2$Sr$_2$Ca$_{n-1}$Cu$_n$O$_{2n+4+x}$ (BSCCO) and YBa$_2$Cu$_3$O$_{7-x}$ (YBCO) exhibit some of the highest known superconducting transition temperatures $T_c$~\cite{RevModPhys.60.585,plakida2010high-temperature}, potentially holding clues to one of the "holy grails" of condensed matter physics – room-temperature superconductivity (sc) at ambient pressure. Superconductivity emerges in cuprates upon hole or electron doping of antiferromagnetic (AF) Mott insulator parent compounds~\cite{RevModPhys.40.677,RevModPhys.66.261}, and is thought to be mediated by the residual AF Mott correlations rather than the conventional Bardeen-Cooper-Schrieffer (BCS) mechanism~\cite{doi:10.1126/science.235.4793.1196,PhysRevB.35.8865,Ogata_2008,Ogata_2008,plakida2010high-temperature,pavarini2017the,RevModPhys.40.677,RevModPhys.66.261}. However, after 35 years of intense research a generally accepted microscopic theory of sc in HTSCs has not yet emerged.

Here, we re-visit one of the earliest models of Mott correlations in doped cuprates, the mean-field (MF) resonating valence bond (RVB) approximation proposed by Baskaran, Zou, and Anderson (BZA) in 1987~\cite{BASKARAN1987973}, and show that it can account for the essential electronic properties of doped cuprates, including the non-Fermi liquid ''strange" normal metal properties, the d wave-like dependence of the paired electrons' phase on the in-plane direction, and the commonly observed charge density modulations. These properties emerge due to the intricate interplay between spin and charge degrees of freedom that become separated in the MF-RVB state.

In Section~\ref{sec:RVB}, we provide an overview of the MF-RVB model of BZA, with additional details and clarifications on the relevance of its predictions to the experimental data that became available in the 35 years since this model was proposed. In Section~\ref{sec:justification}, we examine the assumptions underlying the MF-RVB approximation of BZA. In Section~\ref{sec:electronic}, we analyze the electronic properties expected from this model. We summarize our findings in Section~\ref{sec:discussion}.

\section{Background}\label{sec:RVB}

{\bf The Mott-Hubbard model of the CuO$_2$ plane.} The CuO$_2$ planes hosting sc in cuprates consist of a square lattice of Cu atoms with lattice constant $a$, with an oxygen atom positioned between each pair of the nearest Cu neighbors. In the undoped parent compounds, the $2p$ shell of oxygen is completely filled, while the $4s$ shell of Cu is empty and there is one hole per Cu atom in its $3d$ shell~\cite{Guo1988}. Tetragonal symmetry of the Cu environment splits its five $d$-orbitals into a two-fold orbitally degenerate molecular orbital (MO) level formed due to the hybridization of the Cu $d_{xz}$, $d_{yz}$ orbitals with oxygen's $p$-orbitals, and three orbitally non-degenerate levels derived from the Cu $d_{z^2}$, $d_{xy}$, and $d_{x^2-y^2}$ orbitals~\cite{Guo1988,PhysRevB.37.3759}. Each level is two-fold spin-degenerate. All the orbitals are filled with electrons, except for $d_{x^2-y^2}$, which is half-filled with one hole.

The Mott-Hubbard Hamiltonian for the states derived from the Cu $d_{x^2-y^2}$ MOs is~\cite{doi:10.1126/science.235.4793.1196,pavarini2017the}
\begin{equation}\label{eq:H}
	\hat{H}=\hat{H}_{hop}+\hat{H}_{int}=t\sum_{\vec{n},\vec{l},s}(\hat{c}_{\vec{n}+\vec{l},s}^+\hat{c}_{\vec{n},s})+U\sum_{\vec{n}}\hat{n}_{\vec{n},\uparrow}\hat{n}_{\vec{n},\downarrow}.
\end{equation}
The first term is the kinetic energy described by $t>0$ for Cu-projected hopping, with $\vec{l}$ representing a unit vector in one of the four directions along the principal axes. The second term describes onsite Coulomb interaction, with $U>0$. We utilize the hole representation throughout this work, which facilitates the analysis of the effects of hole doping presented below. In the hole representation, $\hat{c}^+_{\vec{n},s}$ creates a hole in the MO centered on the $\vec{n}^{th}$ Cu site and $\hat{n}_{\vec{n},s}=\hat{c}^+_{\vec{n},s}\hat{c}_{\vec{n},s}$ is the hole number operator. The spin is $s=\pm\frac{1}{2}$, or equivalently $s=\uparrow,\downarrow$.

The effects of kinetic energy in Mott insulators are usually analyzed perturbatively in terms of $t/U$. Projection on the states with one hole per site yields the Hamiltonian~\cite{PhysRevB.2.4302},
\begin{equation}\label{eq:Hex2}
	\hat{H}^{(2)}_{hop}=-J\sum\hat{n}_{\vec{n},s}\hat{n}_{\vec{n}+\vec{l},-s}-J\sum\hat{c}_{\vec{n},s}^+\hat{c}_{\vec{n}+\vec{l},-s}^+\hat{c}_{\vec{n},-s}\hat{c}_{\vec{n}+\vec{l},s},
\end{equation}
where $\hat{n}_{\vec{n}}$ is the particle number operator on site $\vec{n}$, $J=4t^2/U$ is the Heisenberg exchange constant, and summation is performed over all the dummy indices. This equation can be also written as the Heisenberg exchange interaction $\hat{H}_{ex}=J\sum(\vec{\hat{S}}_{\vec{n}}\vec{\hat{S}}_{\vec{n}+\vec{l}}-1/4)/2$~\cite{doi:10.1126/science.235.4793.1196}, where $\vec{\hat{S}}_{\vec{n}}=\hat{c}^+_{\vec{n},s}\vec{\sigma}_{ss'}\hat{c}_{\vec{n},s'}/2$ is the spin operator on site $\vec{n}$, and $\vec{\sigma}$ is the vector of Pauli matrices.

The Schrieffer-Wolf transformation of the Hubbard Hamiltonian Eq.~(\ref{eq:H}) yields an additional hopping ("t") term, which allows one to describe single electron transport not captured by Eq.~(\ref{eq:Hex2})~\cite{Fazekas1999-fn}. This term considered by BZA yields gapped single-particle charge excitations, allowing one to describe single-particle transport. An important conclusion based on the analysis presented below is that the simpler Heisenberg model given by Eq.~(\ref{eq:Hex2}), which cannot describe single-particle charge excitations of the t-J model, is nevertheless capable of describing gapless two-electron spin singlet states. We argue that this result is consistent with the experimental observations of non-Fermi liquid normal metallic state of doped cuprates, providing a possible resolution to the puzzling properties of this metallic state~\cite{plakida2010high-temperature}.  

We limit the discussion of the BZA model to the J-only limit, and leave the full t-J analysis of the effects revealed by our work to future studies. To this end, we point out that the few-site models presented below to justify the mean-field approximation and to demonstrate the mechanism of superconductivity by spinon condensate drag do not rely on the J-only approximation, but are rather based on the full Hubbard Hamiltonian Eq.~(\ref{eq:H}). The agreement between these results and the J-only limit of the BZA model supports our expectation that our main conclusions must remain valid in the full t-J model.

{\bf MF-RVB approximation of BZA.} The Heisenberg Hamiltonian Eq.~(\ref{eq:Hex2}) describes AF interactions between neighboring spins. Accordingly, undoped parent compounds of cuprate superconductors are AF-ordered Mott insulators. Doping destabilizes the AF order, resulting in the emergence of a non-Fermi liquid metal~\cite{plakida2010high-temperature}. BZA described the residual pairwise AF correlations by the singlet order parameter
\begin{equation}\label{eq:correlatorBZA}
	\Delta=\sqrt{2}\langle\hat{b}_{\vec{n},\vec{l}}\rangle,
\end{equation}
where 
\begin{equation}\label{eq:b}
	\hat{b}_{\vec{n},\vec{l}}=\frac{1}{\sqrt{2}}(\hat{c}_{\vec{n}+\vec{l},\downarrow}\hat{c}_{\vec{n},\uparrow}-\hat{c}_{\vec{n}+\vec{l},\uparrow}\hat{c}_{\vec{n},\downarrow})
\end{equation}
is the singlet annihilation operator. This correlator allowed BZA to reduce the Hamiltonian Eq.~(\ref{eq:Hex2}) to a quadratic form, similarly to the Bogolyubov's approach to BCS~\cite{tinkham2004introduction}, as follows. Inserting the identity
\begin{equation}\label{eq:identity}
	\hat{b}_{\vec{n},\vec{l}}=\frac{\Delta}{\sqrt{2}}+[\hat{b}_{\vec{n},\vec{l}}-\frac{\Delta}{\sqrt{2}}]
\end{equation}
and a similar identity for the creation operator into Eq.~(\ref{eq:Hex3}), and keeping only linear in fluctuations terms [terms in parentheses in Eq.~(\ref{eq:identity})], one obtains the linearized MF Hamiltonian
\begin{equation}\label{eq:Hexmf}
	\hat{H}_{mf}=2JN|\Delta|^2-\frac{J\Delta^*}{\sqrt{2}}\sum\hat{b}_{\vec{n},\vec{l}}+h.c.,
\end{equation}
where "h.c." is Hermitian conjugate of the last term, and $N$ is the total number of sites. In the momentum representation, this Hamiltonian is 
\begin{equation}\label{eq:Hk}
	\hat{H}_{mf}=2JN|\Delta|^2-2J\Delta^*\sum[\cos(k_xa)+\cos(k_ya)]\hat{c}_{\vec{k},\downarrow}\hat{c}_{-\vec{k},\uparrow}+h.c.
\end{equation}
BZA considered only real $\Delta$ under the assumption that its phase is averaged out~\cite{BASKARAN1987973}, which is not borne out by the analysis presented below. For complex $\Delta=e^{2i\varphi}|\Delta|$, the Hamiltonian is diagonalized by the canonical transformation
\begin{equation}\label{eq:canonical}
	\hat{c}_{\vec{k},s}=\frac{e^{i\varphi}}{\sqrt{2}}(\gamma_{\vec{k},s}\pm2s\gamma^+_{-\vec{k},-s})
\end{equation}
where $\gamma_{\vec{k},s}$ are the Bogolybovon quasiparticle operators. This transformation is qualitatively different from the canonical transformation in the BCS theory. In BCS, the canonical rotation angle continuously varies with electron-electron interaction strength and temperature $T<T_c$, resulting in a second-order superconducting transition~\cite{tinkham2004introduction,Tilley2019}. In contrast, the transformation Eq.~(\ref{eq:canonical}) does not depend on the values of the model parameters. This result implies that singlet condensation is a first-order transition, allowing for local singlet correlations without long-range phase coherence, as may be expected for 2d U(1)-symmetric systems in the spin liquid vortex state above the Berezinskii-Kosterlitz-Thouless (BKT) transition~\cite{Jose2013-fe}. A similar interpretation of the non-superconducting pseudo-gap regime has been proposed in other analyses~\cite{Anderson2004}.

Both signs in Eq.~(\ref{eq:canonical}) diagonalize the Hamiltonian. Depending on the sign, the resulting calculated quasiparticle mode energies become positive or negative. The latter is unphysical and indicates that the quasiparticle is actually an antiparticle. The Bogolyubovon particles are exchanged with antiparticles by reversing the sign 
in front of the second term in Eq.~(\ref{eq:canonical}), which reverses the sign of the corresponding mode energy. The requirement for the non-negatively defined excitation spectrum is then satisfied by the positive sign in Eq.~(\ref{eq:canonical}) for $|k_x\pm k_y|<\pi/a$, and negative otherwise. BZA expressed this dependence of sign on the wavevector in terms of effective Fermi wavevector defined by $|k_{x}\pm k_{y}|=\pi/a$~\cite{BASKARAN1987973}.

The diagonalized Hamiltonian is
\begin{equation}\label{eq:H_diag}
	\hat{H}_{mf}=2JN(|\Delta|^2-\frac{8|\Delta|}{\pi^2})+2J|\Delta|\sum_{\vec{k},s}|\cos(k_xa)+\cos(k_ya)|\gamma^+_{\vec{k},s}\gamma_{\vec{k},s}.
\end{equation}

To determine the self-consistency condition for $\Delta$, Eq.~(\ref{eq:correlatorBZA}) can be re-written in the momentum representation,
\begin{equation}\label{eq:Delta_k}
	\Delta=\frac{1}{N}\langle\sum_{\vec{k}}[\cos(k_xa)+\cos(k_ya)]\hat{c}_{\vec{k},\downarrow}\hat{c}_{-\vec{k},\uparrow}\rangle.
\end{equation}
In terms of the Bogolyubovon operators, it becomes
\begin{equation}\label{eq:Delta_k2}
	|\Delta|=\frac{4}{\pi^2}-\frac{1}{2N}\langle\sum_{\vec{k}}|\cos(k_xa)+\cos(k_ya)|\gamma^+_{\vec{k},s}\gamma_{\vec{k},s}\rangle.
\end{equation}
This condition is satisfied for any quasiparticle distribution. At $k_BT\gg J$, $\langle\gamma^+_{\vec{k},s}\gamma_{\vec{k},s}\rangle\to1/2$, then $\Delta$ approaches $2/\pi^2$ but does not vanish. Here, $k_B$ is the Boltzmann constant. This result is consistent with the observation of electron pairing and pseudo-gap in the excitation spectrum indicative of a finite $\Delta$ at temperatures $T$ significantly above $T_c$, as further discussed below. This analysis does not include phase fluctuations (Goldstone modes) of $\Delta$, and thus cannot account for the BKT transition, which is expected to define $T_c$, or the additional suppression of coherence due to phase fluctuations at high $T$.

\begin{figure}
	\centering
	\includegraphics[width=0.4\columnwidth]{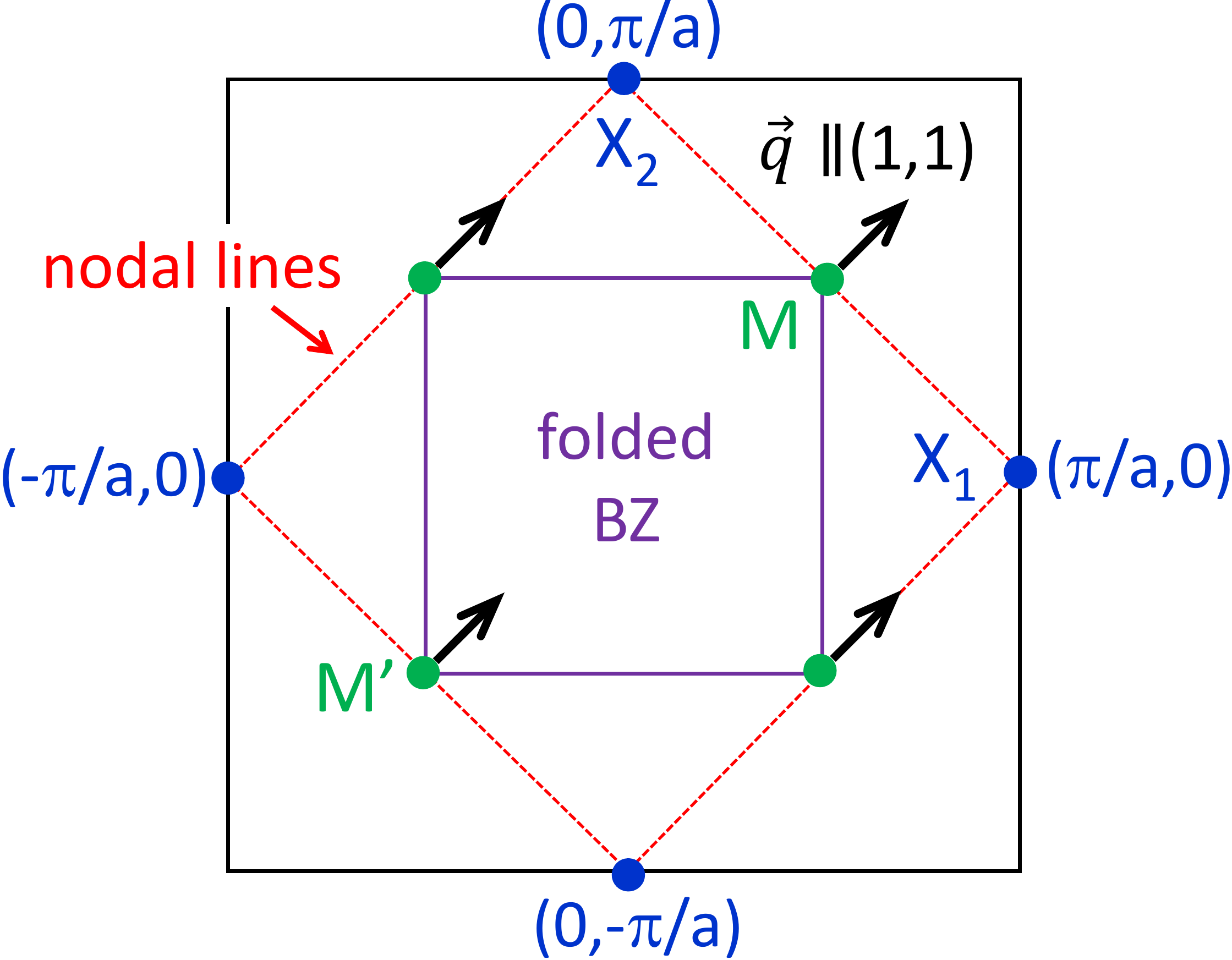}
	\vspace{-5pt}
	\caption{\label{fig:q} Special BZ points, and the effect of order parameter phase gradient along the $(1,1)$ direction at the M-points of the folded BZ.}
\end{figure}

{\bf Fermionic excitations.} BZA identified the Bogolyubovons described by Eq.~(\ref{eq:canonical}) as spinons - chargless spin-$\frac{1}{2}$ fermions, as is also evident if this equation is re-written as
\begin{equation}\label{eq:gamma-c}
	\hat{\gamma}^+_{\vec{k},s}=\frac{e^{-i\varphi}}{\sqrt{2}}(\hat{c}^+_{\vec{k},s}\mp2s\hat{c}_{-\vec{k},-s}),
\end{equation}
where negative sign corresponds to $|k_x\pm k_y|\le\pi/a$. As follows from Eq.~(\ref{eq:H_diag}), the spinon excitation spectrum is
\begin{equation}\label{eq:spectrum}
	\epsilon(\vec{k})=8|\Delta(\cos k_xa+\cos k_ya)|.
\end{equation}
The single-particle excitation spectrum described by Eq.~(\ref{eq:spectrum}) is gapless, with Dirac nodal lines at the wavevectors defined by $|k_x\pm k_y|=\pi/a$ [Fig.~\ref{fig:q}]. BZA also considered the single-electron excitations that emerged due to the hopping term in the t-J model and showed that these excitations are gapped. Such excitations do not appear for the Heisenberg Hamiltonian Eq.~(\ref{eq:Hex2}). Nevertheless, we show below that the latter describes gapless hole pair excitations, consistent with the experimental evidence for non-Fermi liquid normal metal state~\cite{plakida2010high-temperature}. 

\section{Justification for the MF-RVB model of BZA}\label{sec:justification}

Several alternative models have been proposed since the work by BZA, motivated in part by the complexity of experimentally observed properties that appeared to be beyond the scope of this simple model, as discussed in multiple reviews on this subject~\cite{Anderson2004,Berg_2009,PhysRevB.81.054513,RevModPhys.78.17,plakida2010high-temperature}. Here, we analyze the assumptions implicit in this model and re-examine its relevance to the experimental data that became available in the 35 years since it was proposed.

The MF approximation, i.e. the procedure of replacing pairs of field operators with averages, implies that the g.s. is to a good approximation an eigenstate of the operator $\hat{b}_{\vec{n},\vec{l}}$. This should be the case at least in the overdoped limit when singlet correlations are almost completely suppressed and $\Delta$ is small since in this limit $(\Delta-\hat{b}_{\vec{n},\vec{l}})\psi\approx0$. MF approximation then amounts to Taylor expansion of physical quantities such as the Hamiltonian to the lowest order in $\Delta$.

The second assumption implicit in the BZA model is that singlet correlations described by Eq.~(\ref{eq:correlatorBZA}) are independent of the position or the direction between two neighboring sites. This assumption appears to contradict phase-sensitive experiments, which indicate that the sign of the electron pair wavefunction reverses upon rotation by $90^\circ$~\cite{PhysRevLett.74.797}. Indeed, more recent models suggest that d-wave directional symmetry of singlet correlations may be favored~\cite{Zhang_1988,Anderson2004}. Remarkably, the MF-RVB model of BZA describes correlated electron pairs whose wavefunction reverses sign upon $90^\circ$ rotation, as shown in the next section. Here, we examine the assumption that a single parameter $\Delta$ can describe nearest-neighbor singlet correlations. Our analysis consists of two parts. First, we argue that doping reduces the effective size of AF-correlated regions. Second, we consider a minimal RVB model for a small AF-coupled region and show that singlet correlations in its g.s. are direction- and position-independent.

{\bf Effects of doping.} Qualitatively, doping is expected to reduce the effective size of AF-correlated regions because hopping of the dopant holes is suppressed. A dopant hole localized on a certain site rather than being spread into a Bloch wave eliminates spin correlations on this site. Similar conclusions were reached in other analyses based e.g. on the t-J model~\cite{RevModPhys.78.17}.

To the lowest order in $\hat{H}^{(2)}_{hop}$, the g.s. on the subspace of states with a single hole per site is a superposition of ''bare" AF-ordered states $\psi_{AF,\pm}=\prod\hat{c}_{\vec{n},\pm(-1)^{n_x+n_y}/2}^+|0\rangle$, 
\begin{equation}\label{eq:AFstate}
	\psi^{(1)}=\alpha\psi_{AF,+}-\beta\psi_{AF,-},
\end{equation}
where $\alpha$, $\beta$ are coefficients constrained by normalization. The AF-ordered states in undoped cuprates with the N\'eel vector pointing up or down are described by $\alpha=0$, or $\beta=0$, while the singlet state stabilized by doping is described by $\alpha=\beta$.

Consider now the ''bare" AF-ordered state $\psi_{AF,+}$ with an additional spin-down hole, which according to the Pauli principle can have a finite amplitude only on sites $\vec{n}$ with even $n_x+n_y$ 
\begin{equation}\label{eq:1hole}
	\psi=\sum_{n_x+n_y=2n} a_{\vec{n}}\hat{c}_{\vec{n},\downarrow}^+\psi_{AF,+}.
\end{equation}
The total energy of this state is $U$ regardless of the distribution of $a_{\vec{n}}$ that satisfy the normalizing condition $\sum|a_{\vec{n}}|^2=1$, i.e. the hole is effectively localized. This happens because the hopping of spin-down holes from even to odd sites is prohibited by the Pauli principle.

This argument can be extended to the singlet state characterized by $\alpha=\beta$. According to the Pauli principle, the amplitude of the additional spin-down hole is finite for even values of $n_x+n_y$ in the state $\psi_{AF,+}$, and for odd values in the state $\psi_{AF,-}$,
\begin{equation}\label{eq:1hole_ss}
	\psi=\sum_{n_x+n_y=2n}a_{\vec{n}}\hat{c}_{\vec{n},\downarrow}^+\psi_{AF,+}	+\sum_{n_x+n_y=2n+1}a_{\vec{n}}\hat{c}_{\vec{n},\downarrow}^+\psi_{AF,-}.
\end{equation}
For $N>2$, the energy of this state is also $U$, independent of the values of $a_{\vec{n}}$ satisfying the normalizing condition. The result is the same as for the AF-ordered state because the hopping matrix element between the two AF-ordered components vanishes. 

These arguments suggest that dopant holes are quasi-localized, consistent with the small effective hopping parameter expected from the t-J model for hole doping in the hole representation~\cite{PhysRevB.39.6880}. The doped state can then be approximated as a superposition of different distributions of AF-coupled regions with one hole per site separated by doubly occupied sites. The latter do not participate in AF correlations, effectively bounding the AF-coupled regions. The effective size $N$ of AF-coupled regions decreases with increasing doping, justifying the few-site RVB model discussed next.

\begin{figure}
	\centering
	\includegraphics[width=0.3\columnwidth]{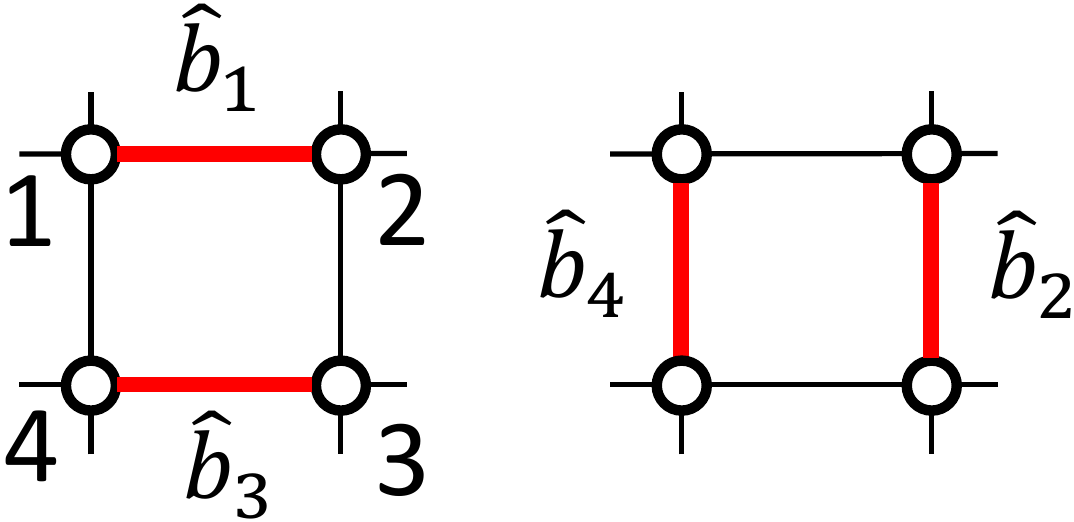}
	%%	\vspace{-5pt}
	\caption{\label{fig:BZ_tiling} Enumeration of sites and nearest-neighbor valence bonds in the RVB state for the model system that consists of four sites on a square. 
	}
\end{figure}

{\bf Four-site RVB model.} Based on the above analysis, the MF-RVB approximation of BZA may be most applicable to heavily doped systems characterized by small AF-correlated regions. Thus, we analyze the symmetry of the MF-RVB order parameter by considering a small model system that consists of four sites arranged on a square, Fig.~\ref{fig:BZ_tiling}. The four possible nearest-neighbor singlet bonds described by operators $\hat{b}_{n}$ can form two pairwise arrangements, as shown in Fig.~\ref{fig:BZ_tiling}, which also illustrates the site and bond enumerations. We use a trial variational RVB wavefunction 
\begin{equation}\label{eq:RVB_trial}
	\psi_{RVB}=(\sum u_m\hat{b}_{m}+v_x\hat{b}_{1}\hat{b}_{3}+v_y\hat{b}_{2}\hat{b}_{4})|1\rangle,
\end{equation}
where $|1\rangle$ is the state in which all sites are filled with two holes. Using the identities

$$\langle1|\hat{b}^+_1\hat{b}^+_3\hat{b}_2\hat{b}_4|1\rangle=\langle1|\hat{b}^+_2\hat{b}^+_4\hat{b}_1\hat{b}_3|1\rangle=1/2$$

obtained from the algebraic relations for singlet operators summarized in the Appendix, we obtain the normalization condition $|v_x|^2+|v_y|^2+Re(v_xv^*_y)=1-4d$, where $d=\sum|u_m|^2/4$ is doping.

Writing the Heisenberg Hamiltonian Eq.~(\ref{eq:Hex2}) as
\begin{equation}\label{eq:Hex3}
	\hat{H}^{(2)}_{hop}=-J\sum_{\vec{n},\vec{l}}\hat{b}^+_{\vec{n},\vec{l}}\hat{b}_{\vec{n},\vec{l}}
\end{equation}
and using the algebraic properties of singlet operators summarized in the Appendix, we obtain the energy of the state Eq.~(\ref{eq:RVB_trial})
\begin{equation}\label{eq:E_RVB}
	E_{RVB}=-3JRe(v_xv^*_y)+C,
\end{equation}
where $C$ is a constant dependent only on the doping level $d$. This energy is minimized by $v_y=v_x\equiv v$, i.e. double-bond amplitudes are independent of bond direction. This analysis does not place any constraints on $u_m$, as long as they satisfy the constraint $n=\sum|u_m|^2/4$ imposed by doping. These coefficients describe the spatial properties of the dopant holes, which are determined by the kinetic energy contributions not included in the Heisenberg Hamiltonian. 

To determine the relation among $u_m$, we consider a simple limiting case of two dopant holes on four sites. Since the interaction term $\hat{H}_{int}$ in Eq.~(\ref{eq:H}) suppresses charge fluctuations, the wavefunction can be approximated as a superposition of states containing one singlet,
\begin{equation}\label{eq:RVB_trial2}
	\psi'_{RVB}=(\sum u_m\hat{b}_{m}+u_{13}\hat{b}_{13}+u_{24}\hat{b}_{24})|1\rangle,
\end{equation}
where the operators
\begin{equation}\label{eq:diag}
	\hat{b}_{13}=\frac{1}{\sqrt{2}}(\hat{c}_{1,\downarrow}\hat{c}_{3,\uparrow}-\hat{c}_{1,\uparrow}\hat{c}_{3,\downarrow}), \hat{b}_{24}=\frac{1}{\sqrt{2}}(\hat{c}_{2,\downarrow}\hat{c}_{4,\uparrow}-\hat{c}_{2,\uparrow}\hat{c}_{4,\downarrow})
\end{equation}
annihilate "diagonal valence bonds" whose relevance to the RVB state becomes apparent from the dependence of energy on the diagonal bond amplitudes.

All the terms in Eq.~(\ref{eq:RVB_trial2}) are characterized by the same interaction energy $\langle \hat{H}_{int}\rangle=2U$. The calculation of the kinetic energy is facilitated by the algebraic relations
\begin{equation}\label{eq:algebra}
	\begin{split}
		\hat{H}_{hop}\hat{b}_{13}|1\rangle=&\hat{H}_{hop}\hat{b}_{23}|1\rangle=-t\sum\hat{b}_m|1\rangle,\\
		\hat{H}_{hop}\hat{b}_m|1\rangle=&-t(\hat{c}_{m,\downarrow}\hat{c}_{m,\uparrow}
		+\hat{c}_{m+1,\downarrow}\hat{c}_{m+1,\uparrow}+\hat{b}_{13}+\hat{b}_{24})|1\rangle,
	\end{split}
\end{equation}
where index $m$ is treated as cyclic, i.e. $m+1\equiv1$ for $m=4$. The kinetic energy is
\begin{equation}\label{eq:E_doped}
	E'_{hop}=\langle\psi|\hat{H}_{hop}|\psi\rangle=-2tRe[\sum u^*_m(u_{13}+u_{23})].
\end{equation}
With the normalization condition $\sum|u_m|^2+|u_{13}|^2+|u_{24}|^2=1$, this energy is minimized if all $u_m$ are equal, $u_m=u$, i.e. single valence bond amplitudes are independent of position. Plugging the wavefunction Eq.~(\ref{eq:RVB_trial}) into the definition Eq.~(\ref{eq:correlatorBZA}) of the singlet order parameter, we obtain $\Delta=\sqrt{2}uv^*$, i.e. the singlet correlator is independent of the position or direction, consistent with the MF approximation Eq.~(\ref{eq:correlatorBZA}) proposed by BZA~\cite{BASKARAN1987973}.

The minimization of the RVB state energy defines the magnitudes of $u$ and $v$ but not their relative phases, resulting in gauge symmetry of the order parameter $\Delta$. BZA contended that the phase of the singlet order parameter is meaningless because it becomes averaged over different valence bond realizations~\cite{BASKARAN1987973}, and others have similarly termed this phase "illusory"~\cite{Zhang_1988}. However, our calculations for a $2\times2$ model system suggest that energy is minimized if the complex coefficients $u$, $v$ describing different contributions to the RVB state are spatially homogeneous, resulting in a well-defined, albeit arbitrary, phase of $\Delta$.

To summarize this section, analysis of the assumptions underlying the MF-RVB model of BZA suggests that it may provide a good approximation for cuprates at least in the heavy-doping limit when the AF-correlated regions become small and valence bond hopping mediated by dopant charges results in direction- and position-independent singlet correlations.

\section{Electronic properties of the MF-RVB state}\label{sec:electronic} 

{\bf Charge excitations.} In the t-J approximation discussed by BZA, gapped single-charge excitations appear as a consequence of a finite t-term. The Heisenberg limit ($t=0$) of this model cannot describe single-charge excitations. Nevertheless, we now show that this approximation is capable of describing gapless singlet hole pair excitations, providing a tentative interpretation, in the framework of this model, for the experimentally observed non-Fermi-liquid normal metal state~\cite{PhysRevB.81.054513}.

Qualitatively, charge excitations are "hidden" in the Heisenberg limit of the MF-RVB model at the intersection of the nodal lines of the spinon spectrum with the BZ boundary, represented by the X-points $\vec{k}_{X}=(0,\pm\pi/a),(\pm\pi/a,0)$ in Fig.~\ref{fig:q}. The X-points on two opposite sides are equivalent. A spinon with momentum $\vec{k}=\vec{k}_{X}-\delta\vec{k}$ close to the X-point is an equal-amplitude superposition of a hole with momentum $\vec{k}$ and an electron with momentum $-\vec{k}$ close to the X-point at the opposite side of BZ. On the other hand, the sign in the canonical transformation Eq.~(\ref{eq:canonical}) is reversed at $|k_x\pm k_y|=\pi/a$, exchanging particles with antiparticles. Both possible signs in the canonical transformation Eq.~(\ref{eq:canonical}) give zero energy of the corresponding quasiparticle mode, indicating that the distinction between particles and antiparticles is lost, making it possible to describe charge excitations. 

To formalize this intuition, we consider a singlet of two spinons with the same wavevector $\vec{k}=\vec{k}_X$ at one of the X-points added to the spinon vacuum $|0_\gamma\rangle$,
\begin{equation}\label{eq:gs_2hs2}
	\psi_{2hs}=\hat{\gamma}^+_{\vec{k},\uparrow}\hat{\gamma}^+_{\vec{k},\downarrow}|0_\gamma\rangle.
\end{equation}
Using the expression Eq.~(\ref{eq:gamma-c}) for spinons in terms of charged particles and the equivalence of $\vec{k}_X$ and $-\vec{k}_X$, this expression can be transformed into
\begin{equation}\label{eq:gs_2hs}
	\psi_{2hs}=\hat{c}^+_{\vec{k},\uparrow}\hat{c}^+_{\vec{k},\downarrow}|0_\gamma\rangle,
\end{equation}
Thus, a singlet of two spinons at the X-point can be equivalently described as a singlet of two holes at the same X-point. Since there are two non-equivalent X-points, this model can accommodate up to four dopant holes. Additional charges can be accounted for by breaking the system up into smaller parts, each accommodating four holes. These holes do not form a Fermi surface, consistent with the low charge mobility expected from the t-J model~\cite{PhysRevB.39.6880}.

The ability of the model to describe gapless charge excitations as singlet hole pairs is not surprising, given that the order parameter describes singlet hole pair amplitude. In BCS, the existence of the Cooper pair condensate allows one to interpret an electron as a superposition of a hole and a Cooper pair, explaining, for example, the Andreev reflection~\cite{tinkham2004introduction}. In the MF-RVB model, the order parameter describes a similar singlet of two holes. Two holes forming a singlet can be added to the system, without energy cost, by bringing them out of the condensate. We conclude that the Heisenberg limit of the MF-RVB model of BZA describes a spin-charge separated metal similar to the Luttinger liquid in 1d~\cite{PhysRevLett.64.1839,Mastropietro2013}.

{\bf Phase gradient of the order parameter.} In BCS, a gradient $\nabla\varphi=\vec{q}$ of the phase of the gauge-symmetric order parameter $\Delta=|\Delta|e^{2i\varphi}$ describes a shift of the momentum of each electron in a Cooper pair near the Fermi surface by $\vec{q}$, resulting in a charge supercurrent~\cite{tinkham2004introduction,Tilley2019}. In contrast, the MF-RVB order parameter describes singlet correlations of Heisenberg spins, which cannot be interpreted in terms of charges on the Fermi surface. Its phase gradient describes the superfluidity of singlet correlations, which is not necessarily associated with a charge supercurrent. Since the condensate can be described as spinon pairs, this flow can be also interpreted as a superflow of chargeless spinon condensate. This has led BZA to dismiss $\Delta$ as a candidate for the superconducting order parameter~\cite{BASKARAN1987973}. We now examine the effects of the phase gradient of MF-RVB order parameter and show that it can describe a supercurrent as charge drag by the spinon superflow. This mechanism is analogous to the spin-charge locking proposed by P.W. Anderson in the more general context of RVB states~\cite{PhysRevLett.96.017001}.

\begin{figure}
	\centering
	\includegraphics[width=0.35\columnwidth]{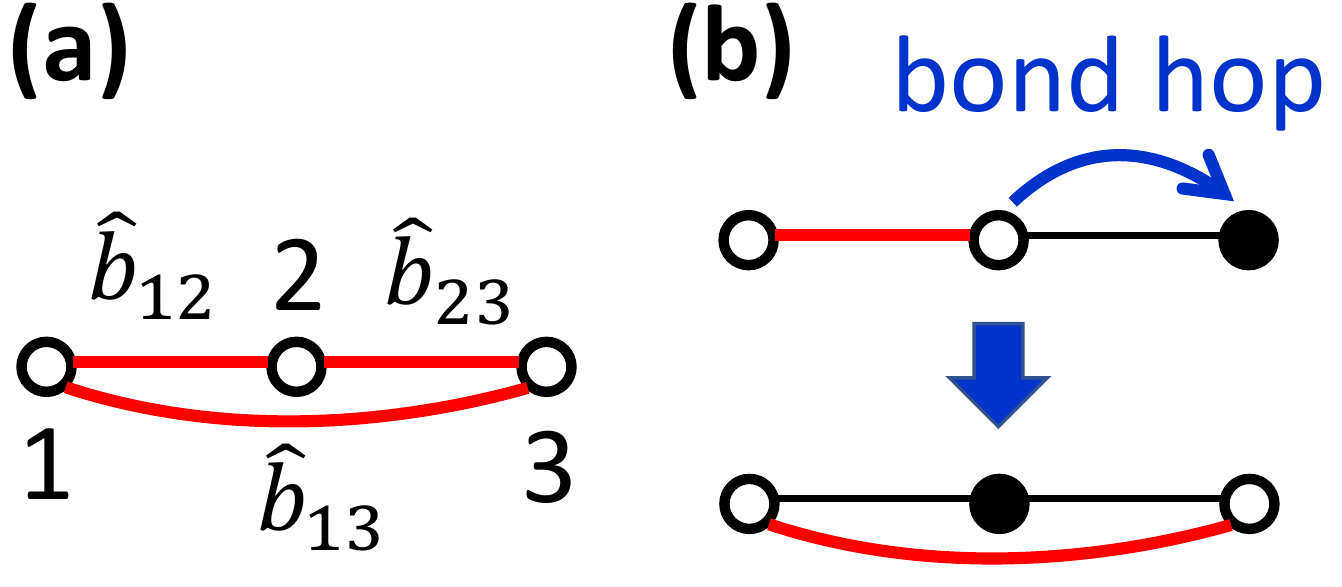}
	\vspace{-5pt}
	\caption{\label{fig:chain} (a) Schematic of the three-site RVB model used to illustrate the effects of order parameter phase gradient. (b) Illustration of charge drag due to valence bond hopping.}
\end{figure}

{\bf Three-site RVB model of phase gradient.} We first consider a minimal RVB model that includes three sites $m=1$, $2$, $3$ on a 1d chain on the x-axis, as illustrated in Fig.~\ref{fig:chain}(a) which also shows the bond enumeration. Consider a trial RVB wavefunction
\begin{equation}\label{eq:RVB_trial3}
	\psi_{RVB}=(u+v_{12}\hat{b}_{12}+v_{23}\hat{b}_{23}+v_{13}\hat{b}_{13})|1\rangle,
\end{equation}
where $\sum|v_m|^2=3(1-d)/2$. The MF-RVB order parameter is $\Delta_1=\sqrt{2}uv^*_{12}$ between sites 1 and 2, and $\Delta_2=\sqrt{2}uv^*_{23}$ between sites 2 and 3. The inclusion of the valence bond connecting sites 1 and 3 to the RVB state is justified by the fact that the hopping Hamiltonian has finite matrix elements between this valence bond and the nearest-neighbor valence bonds, i.e. the hopping Hamiltonian describes bond hopping - hopping of one of the bond ends to the nearest-neighbor site. 

The energy of the state Eq.~(\ref{eq:RVB_trial3}) is 
\begin{equation}\label{eq:E_RVB2}
	E_{RVB}=-2tRe[v^*_{13}(v_{12}+v_{23})].
\end{equation}
In the absence of a phase gradient, energy is minimized by $v_{12}$=$v_{13}=v_{23}/\sqrt{2}=v$, with complex $v$. Phase gradient $d\Delta/dx=2q$ along the chain entails $v_{12}=e^{iqa}v$, $v_{23}=e^{-iqa}v$, then energy is minimized by $v_{13}=\sqrt{2}v$.

The charge current carried by this state is 
\begin{equation}\label{eq:I}
	I=\frac{e}{i\hbar}\langle\psi_{RVB}|[\hat{H}_{hop},\hat{x}/a]|\psi_{RVB}\rangle=\frac{4te|v|^2}{\hbar}\sin qa,
\end{equation}
where $\hat{x}=a\sum_{m,s}m\hat{n}_{m,s}$ is the hole coordinate operator. This current originates from the dependence of charge distribution on the valence bond configuration, which results in the drag of the dopant hole by the singlet flow. For instance, bond $\hat{b}_{12}$ forces the dopant hole onto site 3, while bond $\hat{b}_{13}$ forces the dopant hole onto site 2. Thus, bond hopping from $\hat{b}_{12}$ to $\hat{b}_{13}$ results in hole hopping to the left [Fig.~\ref{fig:chain}(b)], while the reverse process results in hole hopping to the right. Finite $q$ breaks the symmetry between these processes, resulting in a directional hole flow. The symmetry between hopping from bond $\hat{b}_{23}$ to $\hat{b}_{13}$ and the reverse process is also broken, adding to the first contribution.

{\bf Phase gradient in MF-RVB approximation.} The same effect of charge drag by singlet superflow appears in the MF-RVB model of BZA. In the presence of a phase gradient, the canonical transformation becomes
\begin{equation}\label{eq:spinon_q}
	\hat{\gamma}^+_{\vec{k},s}=\frac{e^{-i\varphi}}{\sqrt{2}}(\hat{c}^+_{\vec{k}+\vec{q},s}\mp2s\hat{c}_{-\vec{k}+\vec{q},-s}),
\end{equation}
which shows that phase gradient shifts the momenta of both the hole and the electron components of spinons by $\vec{q}$, similar to the BCS result for Bogolyubovons~\cite{tinkham2004introduction,Tilley2019}. 

To determine the current carried by the dopant holes, we first calculate the hole velocity operator for the hopping Hamiltonian Eq.~(\ref{eq:H})
\begin{equation}\label{eq:velocity}
	\hat{v}=\frac{1}{i\hbar}[\hat{H}_{hop},\vec{\hat{r}}]=-\frac{2ta}{\hbar}\sum(\sin k_xa,\sin k_ya)\hat{c}^+_{\vec{k},s}\hat{c}_{\vec{k},s},
\end{equation}
where $\vec{\hat{r}}=a\sum\vec{n}\hat{n}_{\vec{n},s}$ is the position operator. The corresponding sheet current density carried by a hole with wavevector $\vec{k}$ is 
\begin{equation}\label{eq:jk}
	\vec{j}_{\vec{k}}=\frac{e}{Na^2}\langle\vec{\hat{v}}\rangle=-\frac{2te}{\hbar aN}(\sin k_xa,\sin k_ya).
\end{equation}

According to Eq.~(\ref{eq:spinon_q}), the dopant holes that correspond to the "phase-twisted" spinons at the X-points are characterized by the wavevectors $(\pm\pi/a,0)+\vec{q}$ and $(0,\pm\pi/a)+\vec{q}$. The total sheet current density carried by these holes is then 

\begin{equation}\label{eq:jq}
	\vec{j}(\vec{q})=\frac{tde}{\hbar a}(\sin q_xa,\sin q_ya).
\end{equation}
This result is consistent with Eq.~(\ref{eq:I}) for the three-site RVB chain discussed above. For an order of magnitude estimate, we use typical values $t=0.5$~eV, $a=0.4$~nm, $d=0.1$, to obtain the maximum sheet current density $j_{max}=3\times10^4$A/m allowed by this mechanism, corresponding to the maximum volume current density of the order $10^{13}$A/m$^2$, about four orders of magnitude larger than typical experimental values~\cite{plakida2010high-temperature}. The model does not include phase fluctuations or the effects of spatial inhomogeneity that can result e.g. in phase slips, preventing direct comparison with experimental observations. 

{\bf Symmetry of dopant wavefunctions.} Naively, the direction- and position-independent order parameter $\Delta$ in the MF-RVB model of BZA appears to be inconsistent with the experiments indicating that the phase of superconducting electron pairs reverses upon rotation by $90^\circ$~\cite{PhysRevLett.74.797}, or with observations of spatial charge density modulation~\cite{Norman1998,Hoffman2002,plakida2010high-temperature}. Here, we show that the MF-RVB state that includes both the singlet condensate and the dopant holes is consistent with these experiments.

If the four-fold symmetry of the CuO$_2$ plane is not broken, the same hole populations are expected for both inequivalent X-points of BZ. To determine the relationship between the wavefunctions of the holes localized at these points, we consider a superposition of these two-hole states
\begin{equation}\label{eq:dwave_1}
	\psi=\alpha\psi_{X1}+\beta\psi_{X2},
\end{equation}
where $\alpha$, $\beta$ are coefficients constrained by normalization, and $\psi_{X1}$, $\psi_{X2}$ are the two-hole wavefunctions of holes localized at the X-points $(\pi/a,0)$ and $(0,\pi/a)$, correspondingly. In the coordinate representation, these states are 
\begin{equation}\label{eq:psi_X}
	\psi_{X1}=\frac{1}{N}\sum (-1)^{n_{1x}+n_{2x}}\hat{c}^+_{\vec{n}_1,\uparrow}\hat{c}^+_{\vec{n}_2,\downarrow}|0_\gamma\rangle,\ 
	\psi_{X2}=\frac{1}{N}\sum (-1)^{n_{1y}+n_{2y}}\hat{c}^+_{\vec{n}_1,\uparrow}\hat{c}^+_{\vec{n}_2,\downarrow}|0_\gamma\rangle.
\end{equation}
In the linear approximation described by the MF Hamiltonian Eq.~(\ref{eq:Hk}), the energy of the state Eq.~(\ref{eq:dwave_1}) vanishes regardless of the values of $\alpha$, $\beta$, as expected for the nodal lines of the spinon spectrum. However, the nonlinear exchange energy of this state described by the Heisenberg Hamiltonian Eq.~(\ref{eq:Hex2}) is finite,
\begin{equation}\label{eq:interfere}
	\langle\psi|\hat{H}^{(2)}_{hop}|\psi\rangle=-\frac{2J}{N}[|\alpha|^2+|\beta|^2-2Re(\alpha\beta^*)].
\end{equation}
This energy is associated with the interaction between single-particle states, as is evidenced by its inverse dependence on the system size. It is minimized when the amplitudes of the two components are equal in magnitude and opposite in sign. This dependence comes about because, for each component, the exchange energy is minimized only for one direction defined by the wavevectors of the two holes. For the superposition of the two states characterized by two orthogonal wavevector directions, this energy can be reduced due to the interference that lowers the exchange energy simultaneously in both directions.

The wavefunction Eq.~(\ref{eq:dwave_1}) with $\beta=-\alpha$ is antisymmetric with respect to rotation by $90^\circ$, consistent with the phase-sensitive measurements such as double Josephson junction experiments~\cite{PhysRevLett.74.797}. In the MF-RVB model, this symmetry is associated not with the order parameter $\Delta$, which is direction-independent, but with the two-hole wavefunction of dopants that carry supercurrent.

{\bf Charge density modulation.} We qualitatively discuss how charge density modulations can emerge due to the interplay between hopping-driven two-hole correlations discussed above and the Coulomb interactions. Quantitative analysis is beyond the scope of this work. 

The g.s. Eq.~(\ref{eq:dwave_1}) with $\beta=-\alpha$ describes two dopant holes forming a spin singlet equally distributed between the two inequivalent X-points. As discussed above, the model can accommodate additional holes by breaking the system up into smaller subsystems. This may explain the $4a\times4a$ charge density modulation that becomes particularly prominent around doping $d\approx 1/8$~\cite{Norman1998,Hoffman2002,plakida2010high-temperature,Yoshida_2007}. At this doping, two holes are distributed over $16$ unit cells of the CuO$_2$ plane, i.e. the highest-symmetry system that can accommodate the singlet Eq.~(\ref{eq:dwave_1}) is a $4a\times4a$ square plaquette. The charge density modulation with the same periodicity can be then attributed to correlations between singlets in the neighboring plaquettes, allowing them to reduce their Coulomb interaction energy by minimizing the spatial overlap of their wavefunctions. This pattern can be alternatively interpreted as Wigner-like crystallization of singlet hole pairs, as suggested by P. W. Anderson ~\cite{Anderson_Wigner}.

We now discuss other possible charge modulations that can be accommodated by the model, and supported in part by experimental observations. As discussed above, dopant charges can be described self-consistently as singlet hole pairs at the intersection of the nodal lines of the spinon spectrum with the boundaries of BZ. For the 2d BZ of the square CuO$_2$ lattice with periodicity $a$, these are the X-points located in the middle of the BZ edges, Fig.~\ref{fig:q}. However, if this symmetry is spontaneously broken by the dopant charge density modulations such as charge density waves (CDWs), the BZ can become transformed in such a way that other points on the nodal lines of the spinon spectrum can host dopant holes. 

The self-consistency condition for such instabilities in the context of the Fermi surface nesting is that the wavevector $\vec{Q}$ of the CDW spans parallel arcs of the Fermi surface~\cite{PhysRevB.77.165135}. For the spinon spectrum nodal lines, this is possible for $\vec{Q}=(\pm\pi/a,\pm\pi/a)$, which correspond to commensurate diagonal CDWs with period $2a$ in x- and y-directions. In the presence of such CDWs, two of the reduced BZ boundaries coincide with the spinon nodal lines, allowing singlet hole pairs to occupy these lines. Further BZ reduction can occur due to two coexisting diagonal CDWs, resulting in the entire BZ boundary coinciding with the nodal lines of the spinon spectrum. 

Further symmetry reduction due to such charge density modulations is characterized by the lattice constant doubling, and consequently, BZ folding as shown in Fig.~\ref{fig:q}. The nodal lines touch the M-points of the folded BZ - the corner points characterized by $\vec{k}_M=(\pm\pi/2a,\pm\pi/2a)$, allowing for the dopant hole pairs to localize at these points, via the same mechanism as discussed above for the X-points of unfolded BZ.

We now show that the BZ folding and hole localization at the M-points is a self-consistent process, in the sense that the latter generally results in lattice constant doubling. The spatial dependence of the wavefunction of two holes residing at the M-points of folded BZ can be described by the superposition 
\begin{equation}\label{eq:folding}
	\psi=\alpha+\sum_m\beta_me^{i\pi(\sigma_1x+\sigma_2y)/a},
\end{equation}
where $\alpha$ is the (dominant) amplitude of the opposite-wavevector component, and $\beta_m$ are the amplitudes of the components characterized by the two-hole wavevectors $k_2=(\sigma_1\pi/a,\sigma_2\pi/a$), where $\sigma_1,\sigma_2=-1,0,1$.

Spatial inhomogeneities stabilize the form of the wavefunction Eq.~(\ref{eq:folding}) that either maximizes or minimizes the charge density at some location, as well as its variation in a particular direction. For example, for $\alpha=\beta_1=1/\sqrt{2}$ and $\beta_{m>1}=0$, where $\beta_1$ is the amplitude of the component with $k_2=(\pi/a,\pi/a)$, the amplitude of the wavefunction is maximized along $x+y=2ma$, and minimized along $x+y=(2m+1)a$, i.e. this state describes a commensurate diagonal CDW with wavevector $\vec{Q}=(\pi/a,\pi/a)$. Similar pair density waves have been proposed in other models of cuprates~\cite{Berg_2009}. Inhomogeneities at different locations can also favor other CDWs described by the corresponding finite $\beta_m$, resulting, on average, in similar amplitudes of the coefficients $\beta_m=\beta e^{i\delta_m}$, and producing a charge density modulation with the period $2a$ in both principal directions. In the reciprocal space, this modulation entails BZ folding as shown in Fig.~\ref{fig:q}.

\begin{figure}
	\centering
	\includegraphics[width=0.55\columnwidth]{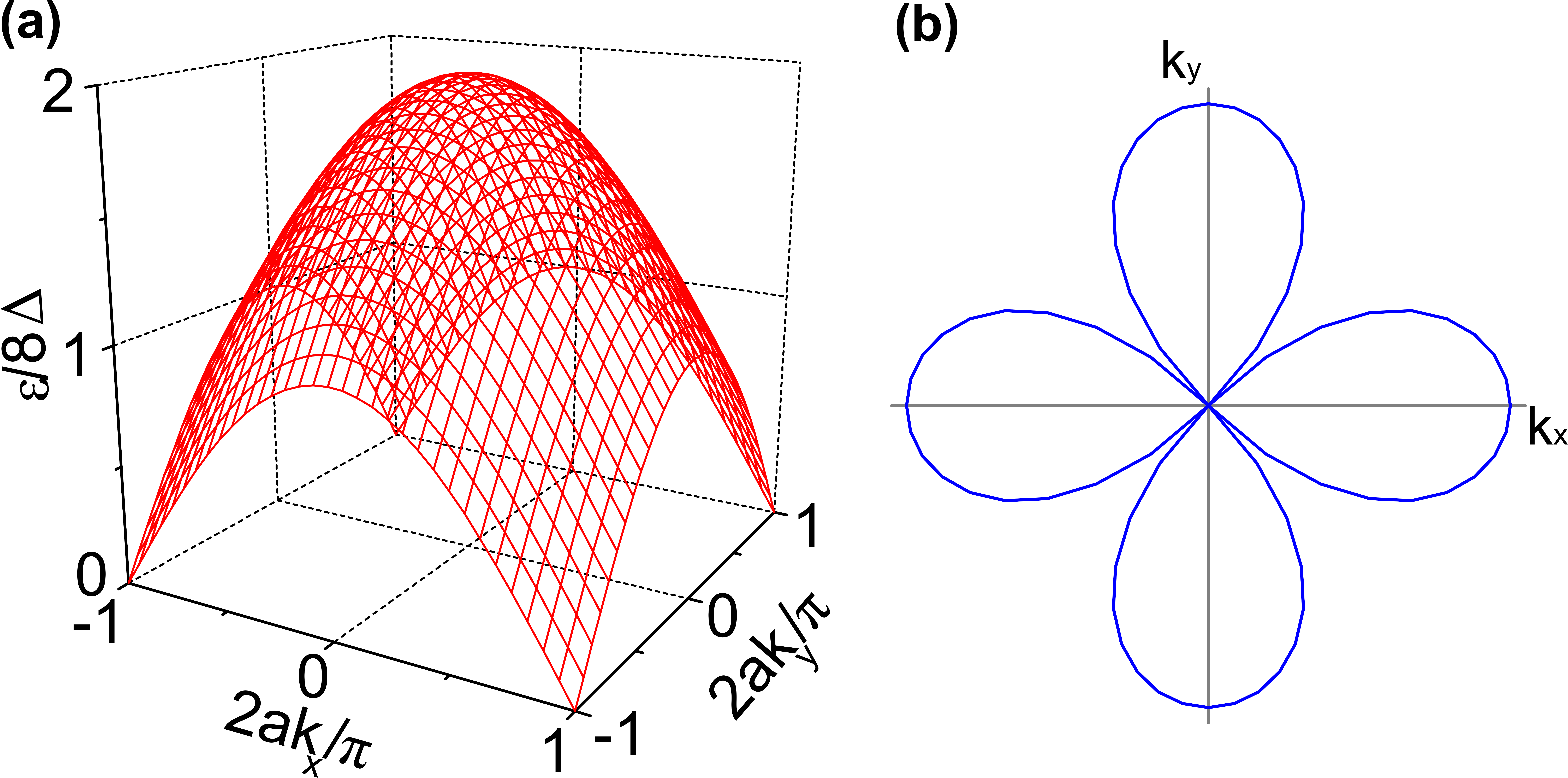}
	\vspace{-5pt}
	\caption{\label{fig:sw_spectrum} (a) Spinon spectrum in the MF-RVB state on the BZ folded by charge ordering. (b) Directional dependence of spinon excitation gap on the folded BZ.}
\end{figure}

The spectrum Eq.~(\ref{eq:spectrum}) defined on the folded BZ is gapless, with four quadrants of a Dirac point at the corners of the folded BZ, Fig.~\ref{fig:sw_spectrum}(a). The directional dependence of the excitation gap in the folded BZ is shown in Fig.~\ref{fig:sw_spectrum}(b). The gap anisotropy is reminiscent of the $d_{x^2-y^2}$ orbital, in excellent agreement with the properties of excitation gap observed in some experiments on doped cuprates, and usually interpreted in terms of the d-wave symmetry of electron pairing~\cite{bok2002the}. In the MF-RVB model of BZA, the gap anisotropy is unrelated to the pairing symmetry, and is instead associated with the directionality of hole hopping on a square lattice. 

If the CuO$_2$ plane is characterized by the $C_4$ symmetry, all four corners are equivalent, and all four may be expected to be occupied by the same hole populations, However, if the $C_4$ symmetry is lifted by the distortions of the CuO$_2$ plane or by its environment, only two of the diagonal points may be expected to become occupied. This scenario is supported for La$_{2-x}$Sr$_x$CuO$_4$ by ARPES measurements, which demonstrate dopant localization at the M-points of the folded BZ along one of the diagonals~\cite{Yoshida_2007}.

{\bf CDW melting due to supercurrent.} In the presence of supercurrent, the order parameter acquires a phase gradient, resulting in a shift of the spinon wavevectors relative to the charged particles, see Eq.~(\ref{eq:spinon_q}). We now argue that this likely results in the suppression of CDW, and consequently BZ unfolding. We consider the effects of phase gradient at the M-points of the folded BZ, $\vec{k}=(\pm\pi/2a,\pm\pi/2a)$. In the absence of phase gradient, the M-points host singlet pairs of dopant holes, which is possible due to their equivalence allowing one to also describe them as spinon singlets. This result is self-consistent, in the sense that the hole pairs can reside at the M-points thanks to the symmetry-breaking charge density modulation resulting in the equivalence of M-points at the opposite corners of the folded BZ. In the absence of CDW, the BZ becomes unfolded. The M-points then become inequivalent and cannot describe charged particles.

This self-consistency is violated at finite $\vec{q}$, because according to Eq.~(\ref{eq:spinon_q}) spinons at the corners of the folded BZ now describe superpositions of charged particles with wavevectors $\vec{k}=(\pm\pi/2a,\pm\pi/2a)+\vec{q}$ that would form CDWs with wavevectors $\vec{Q}=(\pi/a,\pm\pi/a)\pm\vec{q}$ incommensurate with the lattice, which do not lead to BZ folding. Based on these arguments, we conclude that charge density modulations must generally become suppressed at finite $\vec{q}$, resulting in BZ unfolding, and a shift of the dopant states from the M-points of the folded BZ to the X-points of the unfolded BZ. 

The expected perturbation of CDWs by the phase gradient is direction-dependent. Consider, for instance, $\vec{q}=(q,q)/\sqrt{2}$ directed along the BZ diagonal, as illustrated in Fig.~\ref{fig:q}. The CDW with the wavevector $\vec{Q}$ in the same direction is expected to be suppressed by the gradient because the wavevectors of the holes forming this CDW are shifted away from the zone boundary.
On the other hand, the CDW orthogonal to this direction should be unaffected, because $\vec{q}$ is in the direction of its translation symmetry. Thus, for supercurrents along the diagonal direction, the CDW with the wavevector in the same direction may be expected to become suppressed, which can be verified by direct-space imaging techniques such as scanning tunneling microscopy~\cite{Hoffman2002,plakida2010high-temperature}, or by the reciprocal space techniques such as diffraction or ARPES~\cite{Norman1998,Yoshida_2007}.

\section{Summary}\label{sec:discussion}

In this work, we re-examined the Heisenberg limit of the mean-field resonating valence bond (MF-RVB) model proposed in 1987 by Baskaran, Zou, and Anderson (BZA)~\cite{BASKARAN1987973}. We analyzed the assumptions underlying this model and showed that it is likely relevant at least to the highly doped limit when singlet correlations are almost completely suppressed. This analysis leaves open the question of whether other approximations may be more relevant in the light doping regime but nevertheless provides a simple interpretation for some of the experimental results that have puzzled the scientific community for several decades. We have also shown that the behaviors of few-site model systems described by the full Hubbard Hamiltonian, without the mean-field approximations, are consistent with the analysis based on the Heisenberg limit of the BZA model. These findings suggest that the single electron hopping described by the additional ''t"-term in the t-J model based on the Hubbard Hamiltonian underlying this analysis may provide a quantitative rather than a qualitative correction to the electronic properties dominated by many-electron effects.

In the MF-RVB model, spin singlet correlations are described by the gauge-symmetric singlet condensate akin to the condensate of Cooper pairs in the BCS theory of sc, and the fermionic excitations are chargeless spinons. Our central result based on the analysis of the spinon excitation spectrum is that this model is capable of describing charge excitations as spinless two-hole singlets localized at the intersection between the nodal lines of the spinon spectrum and the Brillouin zone (BZ) boundary. This result provides a tentative interpretation of the non-Fermi liquid metal properties of doped cuprates in a simple mean-field framework. 

The MF-RVB singlets are somewhat akin to the Zhang-Rice singlets, which are also formed by a singlet pair of holes, one localized on a Cu atom, another on its oxygen coordination~\cite{PhysRevB.37.3759}. In contrast to the Zhang-Rice singlets whose energies are gapped by the energy of the oxygen p-orbitals, the MF-RVB singlets are gapless, suggesting that Zhang-Rice singlets likely become dissolved in the MF-RVB state. Given the experimental observations of a continuous evolution of singlet hole pairs with doping~\cite{PhysRevLett.115.027002}, it is likely that the Zhang-Rice singlets become increasingly delocalized with increasing doping, continuously transforming into spatially-delocalized (momentum-localized) singlets predicted by the MF-RVB model.

We showed that the superflow of singlet correlations described by the gradient of the singlet condensate order parameter also carries a charge supercurrent, which can be interpreted as the drag of singlet dopant hole pairs by spinon superflow. The wavefunction of the singlet pairs is antisymmetric with respect to the rotation by $90^\circ$, providing a simple interpretation for the phase-sensitive experiments on cuprates~\cite{PhysRevLett.74.797}.
In contrast to the BCS theory of superconductivity, singlet condensation in the MF-RVB model is a first-order transition. This allows for the existence of local singlet pairs in the absence of phase coherence, accounting for the observations of pseudo-gap~\cite{Timusk1999} and electron pairing in non-superconducting state~\cite{Zhou2019}.

The localization of singlets formed by dopant charges at the edges of the BZ may also account for the charge density modulations commonly observed in doped cuprates, as a consequence of the interplay between hopping-induced correlations and Coulomb interaction maximizing the distance between singlet pairs. These modulations can result in the BZ folding, resulting in the excitation gap that closely resembles the shape of the atomic $d_{x^2-y^2}$ orbital, in agreement with some experimental observations. In the MF-RVB model, this directional dependence originates from the anisotropy of electron hopping in the CuO$_2$ plane and is unrelated to the symmetry of electron pairing.

Finally, we note that the MF-RVB model describes a spin-charge separated metal characterized by a gapless spectrum of spin-$\frac{1}{2}$ chargeless fermions, and a gapless excitation spectrum of spinless charges 2$e$, akin to the Luttinger liquid in 1d~\cite{PhysRevLett.64.1839,Mastropietro2013}, supporting the possibility that the Luttinger liquid state is not limited to 1d~\cite{Anderson2004}. It may be possible to explain unconventional superconductivity in other exotic systems in terms of a similar mechanism of charge drag by spin superflow in the spin-charge separated Luttinger liquid state. 

\section*{Acknowledgments}
The contributions by SU were supported by the NSF Awards ECCS-1804198 and ECCS-2005786. Analysis of the RVB model by SI was supported by the DOE BES Award \# DE-SC0018976.

\section*{Appendix: algebra of singlet operators}\label{sec:algebra}

To facilitate the analysis of the Hamiltonian and the trial RVB states written in terms of singlet operators defined by Eq.~(\ref{eq:b}), we summarize their algebraic properties. In our approach, all the operators and states are defined in the hole basis, with $|1\rangle$ denoting a state completely filled with two holes per site.

\begin{enumerate}
	
	\item Creation or annihilation operators commute,
	
	$[\hat{b}_{\vec{n},\vec{l}},\hat{b}_{\vec{n}',\vec{l}'}]=0$,
	
	\item Singlet creation operator commutes with singlet annihilation operator if the two bonds do not overlap,
	
	$[\hat{b}^+_{\vec{n},\vec{l}}\,,\hat{b}_{\vec{n}',\vec{l}'}]=0$ if $\vec{n}\ne\vec{n}'$, $\vec{n}+\vec{l}\ne\vec{n}'$, $\vec{n}\ne\vec{n}'+\vec{l}'$, and $\vec{n}+\vec{l}\ne\vec{n}'+\vec{l}'$,
	
	\item For the same bond,
	
	$[\hat{b}^+_{\vec{n},\vec{l}}\,,\hat{b}_{\vec{n},\vec{l}}\,]=\sum_s(\hat{n}_{\vec{n},s}+\hat{n}_{\vec{n}+\vec{l},s})/2-1$,
	
	\item For two overlapping bonds,
	
	$[\hat{b}^+_{\vec{n},\vec{l}}\,,\hat{b}_{\vec{n},\vec{l}'}]=\sum_s\hat{c}_{\vec{n}+\vec{l}',s}\hat{c}^+_{\vec{n}+\vec{l},s}/2$
	if $\vec{l}\ne\vec{l}'$,
	
	\item
	$\hat{b}^2_{\vec{n},\vec{l}}\,|1\rangle=\hat{c}_{\vec{n},\uparrow}\hat{c}_{\vec{n}+\vec{l},\uparrow}\hat{c}_{\vec{n},\downarrow}\hat{c}_{\vec{n}+\vec{l},\downarrow}|1\rangle$,
	
	\item $\hat{b}^+_{\vec{n},\vec{l}}\,|1\rangle=\hat{c}^+_{\vec{n},s}|1\rangle=0$.
	
\end{enumerate}

These properties allow one to establish other useful relations, for example $\hat{b}^+_{\vec{n},\vec{l}}\,\hat{b}_{\vec{n},\vec{l}}\,|1\rangle=|1\rangle$ follows from $3$ and $6$, while $\hat{b}^+_{\vec{n},\vec{l}}\,\hat{b}_{\vec{n}',\vec{l}'}\,|1\rangle=0$ for non-identical (overlapping or not) bonds follows from 2, 4 and 6.

\bibliographystyle{apsrev4-2}
\bibliography{HTSC}

\end{document}